\documentclass[a4paper,11pt]{article}
\pdfoutput=1 

\usepackage{jheppub} 

\usepackage[T1]{fontenc} 

\usepackage{tabularx}

\title{\boldmath Minimally modified $A_4$ Altarelli-Feruglio model for neutrino masses and mixings and its experimental consequences}

\author[]{Minjie Lei\footnote{minjielei@umich.edu, ORCID ID: 0000-0002-2679-4609}, }
\author[]{James D. Wells\footnote{jwells@umich.edu, ORCID ID: 0000-0002-8943-5718}}

\affiliation[]{Leinweber Center for Theoretical Physics \\
University of Michigan, Ann Arbor, MI, USA}

\emailAdd{minjilei@umich.edu}
\emailAdd{jwells@umich.edu}

\abstract{We present a simple modification of the Altarelli-Feruglio $A_4$ flavor symmetry model using a minimal number of parameters congruous with the symmetries of the original theory. The resulting model is consistent with all presently known data on neutrino masses and mixings. Furthermore, it makes testable tight predictions for future experiments aimed at improving the measurements of $\sin^2\theta_{12}$, $\sin^2\theta_{13}$, $\delta_{\rm CP}$ and the neutrinoless double-beta decay parameter $|m_{ee}|$. Our model exploits the unique possibility of multiple allowed, yet qualitatively different, contractions of fields charged under the $A_4$ discrete symmetry.}

\begin{document} 

\hspace{1in}
\maketitle
\flushbottom

\section{Introduction}

Although SM has been remarkably consistent with experimental data, it is unable to explain several key issues such as the neutrino mass values and ordering, the number of fermion generations, and the observed values of neutrino mixing parameters. Several approaches have been proposed to determine the theory behind fermion mixing. One of the most promising is discrete flavor symmetry models, under which left- and right-handed fermions, as well as proposed new particles, transform as irreducible representations of some discrete symmetry group. The $A_4$ flavor symmetry model proposed by Altarelli and Feruglio \cite{Altarelli:2010gt} is a particularly interesting approach, being a minimal model which produces tri-bimaximal mixing that was considered at the time a good description of the neutrino mixing matrix. \begin{equation}
\label{eq:alt_U}
U_{TB} = 
\begin{pmatrix}
\sqrt{\frac{2}{3}} \ & \frac{1}{\sqrt{3}} \ & 0\\
-\frac{1}{\sqrt{6}} \ & \frac{1}{\sqrt{3}} \ & -\frac{1}{\sqrt{2}} \\
-\frac{1}{\sqrt{6}} \ & \frac{1}{\sqrt{3}} \ & \frac{1}{\sqrt{2}}
\end{pmatrix}
\end{equation}
However, recent reactor neutrino experiments \cite{Abe:2011fz, Ahn:2012nd, Ling:2013fta} have indicted that the $U_{e3}$ component of the mixing matrix deviate sizably from zero with $|  U_{e3} | \approx 0.15$ in violation of the tri-bimaximal pattern that requires $|U_{e3}|=0$.  Thus, $A_4$ is presently considered a good approximation upon which to build in corrections.  In this paper, by fully considering field contractions under $A_4$ symmetry group, we propose a simple modification to the Altarelli-Feruglio model that produce oscillation parameters that agree well with up-to-date experimental data, and discuss further implications.

\begin{table}[tbp]
\centering
\begin{tabular}{ccccccccccc}
      \hline \multicolumn{1}{c|}{}     & $l$      & $e^c$      & $\mu^c$    & $\tau^c$   & $\nu^c$    & $h_{u,d}$ & $\theta$ & $\varphi_T$ & $\varphi_S$ & $\xi/\tilde{\xi}$      \\ \hline
\multicolumn{1}{c|}{$A_4$}       & 3        & 1          & 1''        & 1'         & 3          & 1       & 1        & 3           & 3           & 1          \\
\multicolumn{1}{c|}{$Z_3$}       & $\omega$ & $\omega^2$ & $\omega^2$ & $\omega^2$ & $\omega^2$ & 1       & 1        & 1           & $\omega^2$  & $\omega^2$ \\
\multicolumn{1}{c|}{$U(1)_{FN}$} & 0        & 4          & 2          & 0          & 0          & 0       & -1       & 0           & 0           & 0          \\
\multicolumn{1}{c|}{$U(1)_R$}    & 1        & 1          & 1          & 1          & 1          & 0       & 0        & 0           & 0           & 0 	           \\ \hline
\end{tabular}
\caption{Field content and symmetry group assignment of Altarelli $A_4$ model \cite{Altarelli:2010gt}.}
\label{tab:assignment}
\end{table}

\section{Altarelli-Feruglio $A_4$ Model}

The field content and the symmetry group assignments of the Altarelli-Feruglio model~\cite{Altarelli:2010gt} are given in Table 1. The leptons are assigned to the four irreducible representations of $A_4$.  A Majorana seesaw realization for neutrino masses arises, with RH neutrino field $\nu^c$ transforming as a triplet of $A_4$. The $A_4$ flavor symmetry is broken by two flavon triplets $\varphi_S$, $\varphi_T$ and by one or more singlets $\xi$. Two Higgs doublets $h_{u, d}$ invariant under $A_4$ are also introduced. A supersymmetric (SUSY) context is adopted, with the superpotential term for lepton masses given by:
\begin{equation}
\label{eq:super}
\begin{split}
\omega_l = \ & \frac{y_e}{\Lambda} (l \varphi_T)_1 h_d e^c +  \frac{y_\mu}{\Lambda} (l \varphi_T)_{1'} h_d \mu^c +  \frac{y_\tau}{\Lambda} (l \varphi_T)_{1''} h_d \tau^c \\
& + y (l h_u \nu^c)_1 + (x_A \xi + \tilde{x_A} \tilde{\xi}) (\nu^c \nu^c)_1 + x_B (\varphi_S (\nu^c \nu^c)_{3_s})_1 + h.c. 
\end{split}
\end{equation}
where $\Lambda$ is the cutoff scale. Additional symmetries $U(1)_{FN}$, $U(1)_R$ are also implemented in the model. $U(1)_{FN}$, broken by the $A_4$ singlet field $\theta$, reproduces the observed hierarchy among $m_e$, $m_\mu$, and $m_\tau$. Its overall contribution to the superpotential charged lepton mass terms are implicitly assumed. $U(1)_R$, broken at low energy scale $m_{SUSY}$ to discrete R-parity, is a common feature of supersymmetric formulations. A supplementary $Z_3$ symmetry is also implemented, restricting additional terms otherwise allowed by $A_4$ symmetry, such as the term obtained by exchanging $\varphi_T \leftrightarrow \varphi_S$. 

In the model setup, it can be derived that the flavon fields naturally develop a vacuum expectation value (VEV) along the directions \cite{Altarelli:2009kr}:
\begin{equation}
\label{eq:vac_align}
\left <\varphi_T \right >  = (\nu_T, 0, 0); \ \ \ \left <\varphi_S \right >  = (\nu_S, \nu_S, \nu_S); \ \ \  \left <\xi \right > = v_{\xi}; \ \ \  \langle \tilde{\xi} \rangle = 0 ;
\end{equation}
With this vacuum alignment realized, the mass matrices of the charged lepton and neutrino sectors can be calculated after flavor and electroweak symmetry breaking. At leading order, working in the $T$-diagonal basis for the  $A_4$ triplet representation, the charged lepton mass matrix is given by:
\begin{equation}
\label{eq:alt_cl}
m_l = \frac{\nu_d \nu_T}{\Lambda}
\begin{pmatrix}
y_e \ & 0 \ & 0\\
0 \ & y_{\mu} \ & 0 \\
0 \ & 0 \ & y_{\tau}
\end{pmatrix}
\end{equation}
where the suppression coming from the breaking of $U(1)_{FN}$ is implicit. 
Similarly in the neutrino sector, the Dirac and Majorana masses after symmetry breaking are:
\begin{equation}
\label{eq:alt_nu}
m_D = y v_u
\begin{pmatrix}
1 \ & 0 \ & 0\\
0 \ & 0 \ & 1 \\
0 \ & 1 \ & 0
\end{pmatrix}, \ \ \ \ \
M_R = 
\begin{pmatrix}
A+2B/3 \ & -B/3 \ & -B/3\\
-B/3 \ & 2B/3 \ & A-B/3 \\
-B/3 \ & A-B/3 \ & 2B/3
\end{pmatrix}
\end{equation}
where 
\begin{equation}
\label{eq:alt_param}
A \equiv x_A v_{\xi}, \ \ \ \ \ B \equiv 3 x_B \nu_S.
\end{equation}
The light neutrino mass matrix is $m_{\nu} = (m_D)^T M_R^{-1} m_D$ with eigenvalues: 
\begin{equation}
\label{eq:alt_mass}
m_1 = \frac{y^2 \nu_u^2}{M_1}, \ \ \ \ \ \ m_2 = \frac{y^2 \nu_u^2}{M_2}, \ \ \ \ \ \ m_3 = \frac{y^2 \nu_u^2}{M_3}
\end{equation}
where $M_1$, $M_2$, $M_3$ are eigenvalues of $M_R$ given by:
\begin{equation}
\label{eq:alt_M}
M_1 = (A+B), \ \ \ M_2 = A, \ \ \ M_3 = (-A+B).
\end{equation}
From the form of the mass eigenvalues, one can derive bounds on the lightest neutrino mass and the possible values of the effective Majorana mass $| m_{ee} |$ \cite{Altarelli:2009kr}. For the normal hierarchy:
\begin{equation}
\label{eq:alt_mbound_nh}
\begin{split}
& m_1 \geq \sqrt{\frac{\Delta  m_{sol}^2}{3}} \left (1-\frac{4\sqrt{3}}{9} r + O(r^2) \right ) \approx 0.004\, {\rm eV}
\\
& m_1 \leq \sqrt{\frac{\Delta  m_{sol}^2}{3}} \left (1+\frac{4\sqrt{3}}{9} r + O(r^2) \right ) \approx 0.006\, {\rm eV}
\\
& | m_{ee} | \approx \frac{4}{3\sqrt{3}} \Delta m_{sol}^2 \approx 0.007\, {\rm eV}
\\
\end{split}
\end{equation}
And for inverse hierarchy:
\begin{equation}
\label{eq:alt_mbound_ih}
\begin{split}
& m_3 \geq \sqrt{\frac{\Delta  m_{atm}^2}{8}} \left (1-\frac{1}{6} r^2 + O(r^3) \right ) \approx 0.017 \, {\rm eV}
\\
& | m_{ee} | \geq \sqrt{\frac{\Delta  m_{atm}^2}{8}} \approx 0.017\, {\rm eV}
\\
\end{split}
\end{equation}
where $\Delta  m_{\rm sol}^2 = \Delta m_{21}^2 = m_2^2 - m_1^2$,  $\Delta  m_{atm}^2 = \Delta m_{31}^2 = m_3^2 - m_1^2$, and $r = \Delta  m_{sol}^2 / \Delta  m_{\rm atm}^2$.

In the basis where the charged lepton matrix is diagonal, the PMNS mixing matrix is just the unitary matrix $U$ that diagonalizes the light neutrino mass matrix $m_{\nu}$:
\begin{equation}
\label{eq:U_diag}
U^{\dagger} m_{\nu} U^* = \text{diag}(m_1, m_2, m_3)
\end{equation}
For $m_{\nu}$ that can be calculated from $m_D$ and $M_R$ given in equation \ref{eq:alt_nu}, $U$ is simply:
\begin{equation}
\label{eq:alt_U}
U = 
\begin{pmatrix}
\sqrt{\frac{2}{3}} \ & \frac{1}{\sqrt{3}} \ & 0\\
-\frac{1}{\sqrt{6}} \ & \frac{1}{\sqrt{3}} \ & -\frac{1}{\sqrt{2}} \\
-\frac{1}{\sqrt{6}} \ & \frac{1}{\sqrt{3}} \ & \frac{1}{\sqrt{2}}
\end{pmatrix}
\end{equation}
which is just the Tri-Bimaximal mixing matrix $U_{TB}$. The Tri-Bimaximal form is in reasonable agreement with observed neutrino mixing matrix and is frequently considered as a good first approximation in many flavor symmetry models. However, as noted above, recent reactor neutrino experiments \cite{Abe:2011fz, Ahn:2012nd, Ling:2013fta} have indicted that the $U_{e3}$  matrix element deviates sizably from zero $|  U_{e3} | \approx 0.15$. Modified models that produce sizable corrections to the mixing parameters have been proposed, but often require more complicated models involving non-vanishing higher order contributions \cite{Hall:2013yha}, soft $A_4$ symmetry breaking terms \cite{Felipe:2013vwa}, misalignment of mass matrix elements \cite{Hollik:2017get}, or larger symmetry groups than $A_4$ (e.g. $T', \Delta(6\cdot10^2), (Z_{18}\times Z_6) \rtimes S_3, T_{13}$) \cite{Holthausen:2012wt, Ahn:2013ema, Perez:2019aqq, Rahat:2018sgs}. Here we propose a minimal modification to the Alteralli $A_4$ model by considering an additional Dirac mass term allowed by $A_4$ symmetry contractions that are also allowed by the other symmetries assumed in the model. The modified model retains the simplicity of the $A_4$ model, and predicts neutrino masses and mixing parameters that are in good agreement with current experimental data.

\section{Modified Altarelli-Feruglio Model}

In the Altarelli-Feruglio $A_4$ model, the Dirac mass is entirely specified by the the standard model Yukawa term $y (l h_u \nu^c)_1$ with no flavon contraction modifications. A natural step to modify the existing model, is to replace the Dirac Yukawa term with terms involving flavons $\xi$ and $\varphi_S$ in parallel to the form of the Majorana terms:
\begin{equation}
\label{eq:mod_term}
\frac{y}{\Lambda} \xi(l h_u \nu^c)_1 + \frac{y'}{\Lambda} (\varphi_S (l h_u \nu^c)_3)_1
\end{equation}
The first term is modified by $\xi$, an $A_4$ singlet. For the second term, the $A_4$ contraction between $l$ and $\nu^c$ is now another triplet that in turn contracts with the triplet $\varphi_S$ field to form an $A_4$ invariant. Both $\xi$ and $\varphi_S$ have zero $U(1)_{FN}$ and $U(1)_R$ charges, preserving these symmetries even after they obtain vevs. The only symmetry that is violated by this modification in the original Altarelli-Feruglio model is the supplementary $Z_3$ symmetry, under which $\varphi_S$ transforms as $\omega^2$, making the second term involving $\varphi_S$ an $\omega^2$ instead of a singlet under the $Z_3$. However, since $Z_3$ is a supplementary symmetry introduced to restrict extra terms otherwise allowed by the $A_4$ symmetry, like the terms obtained by exchanging $\varphi_T \leftrightarrow \varphi_S$ and the term $\nu^c \nu^c$, we can modify it without affecting other elements of the model to allow this modification. This goal is accomplished by making the following change in $Z_3$ field assignment: 
\begin{equation}
\label{eq:mod_sym}
\begin{gathered}
\ \ \ \ l \rightarrow \omega^2
\\
e_R, \mu_R, \tau_R \rightarrow \omega
\end{gathered}
\end{equation}
with all other $Z_3$ field assignments remaining the same. The new term in equation \ref{eq:mod_term} now contracts to a $Z_3$ singlet, and all existing terms are still invariant, while additional terms such as those obtained by exchanging $\varphi_T \leftrightarrow \varphi_S$ and $\nu^c \nu^c$ are still not allowed by this new $Z_3$ symmetry.

As specified by the tensor product rules in eq.~\ref{eq:tensor_prod}, there are two ways to contract two $A_4$ triplets $l$ and $\nu^c$ to another triplet. The form of the projection matrix for each linearly independent contraction is given in sec.~\ref{sec:proj_mat}. Therefore, there could be two independent new terms in the form of eq.~\ref{eq:mod_term}, with different coupling constants. For simplicity of the model, in this paper we will only consider the term with the simpler project matrix:
\begin{equation}
\label{eq:mod_contrac}
\tau^{\rho}_{\mu \nu} = 
\left[
\begin{pmatrix}
0 \ & 0 \ & 0\\
0 \ & 1 \ & 0 \\
0 \ & 0 \ & -1
\end{pmatrix},
\begin{pmatrix}
0 \ & -1 \ & 0\\
0 \ & 0 \ & 0 \\
1 \ & 0 \ & 0
\end{pmatrix},
\begin{pmatrix}
0 \ & 0 \ & 1\\
-1 \ & 0 \ & 0 \\
0 \ & 0 \ & 0
\end{pmatrix}
\right]
\end{equation}

In the rest of the paper, we will show that this simple modification already predicts neutrino masses and oscillation parameters that fit well with current experimental data, and discuss additional implications of the model.

\subsection{Parametrization of the model}

With the additional Dirac triplet contraction term, the charged lepton mass matrix and the Majorana neutrino mass matrix remain the same, and the relationship specified by equation \ref{eq:alt_M} still holds, but the Dirac neutrino mass matrix is now
\begin{equation}
\label{eq:mod_mD}
m_D = 
\begin{pmatrix}
a \ & -b \ & b\\
-b \ & b \ & a \\
b \ & a \ & -b
\end{pmatrix}
\end{equation}
where $a$, $b$ is given by 
\begin{equation}
\label{eq:mod_mD_param}
a = y v_u v_{\xi} / \Lambda; \ \ \ \ b = y' v_u v_S / \Lambda.
\end{equation}
For convenience of analysis, we parametrize the model by defining complex dimensionless parameters $H$ and $\eta$:
\begin{equation}
\label{eq:mod_param}
H = \frac{B}{A}, \ \ \ \ \ \ \eta = \frac{b}{a}
\end{equation}
so that
\begin{equation}
\label{eq:mod_mv}
(m_{\nu})_{ij} = (m_D^T M_R^{-1} m_D)_{ij} = 
k \Sigma_{ij}
\end{equation}
where 
\begin{equation}
\label{eq:mod_k}
k = \frac{y^2 v_u^2 v_{\xi}^2 / \Lambda^2}{x_A v_{\xi} \cdot 3(H^2-1)}
\end{equation}
absorbs the dimension of the problem. As an overall factor in the mass matrix $k$ does not affect the neutrino mixing matrix elements. The phase of $k$ can be absorbed by phase redefinitions of the charged lepton fields, so it can be treated as a real parameter without loss of generality. The components of the matrix $\Sigma$ are given in terms of the complex parameters $H$, $\eta$ by
\begin{equation}
\label{eq:mod_omega}
\begin{split}
& \Sigma_{11} = 6\eta^2 + 6\eta^2 H + 2H + H^2 - 3
\\
& \Sigma_{12} = \Sigma_{21} = -3\eta^2 -3 \eta^2 H + 6 \eta - H + H^2
\\
& \Sigma_{13} = \Sigma_{31} = -3\eta^2 -3 \eta^2 H - 6 \eta - H + H^2
\\
& \Sigma_{22} = -3\eta^2 + 6\eta^2 H  - 6 \eta + 2H + H^2
\\
& \Sigma_{23} = \Sigma_{32} =  6\eta^2 -3 \eta^2 H - H + H^2 - 3
\\
& \Sigma_{33} = -3\eta^2 + 6\eta^2 H  + 6 \eta + 2H + H^2.
\end{split}
\end{equation}

\subsection{Prediction for Neutrino Oscillation Parameters}


The light neutrino mass matrix $m_{\nu}$ is diagonalized by the unitary PMNS matrix $U$ according to eq.~\ref{eq:U_diag}. $U$ can be constructed from direct diagonalization of the hermitian matrix $h = m_{\nu} m_{\nu}^{\dagger}$:
\begin{equation}
\label{eq:mod_diag}
U^{\dagger}h\,U = {\rm diag}(m_1^2, m_2^2, m_3^2).
\end{equation}
Following the framework of calculating oscillation parameters of a generalized neutrino mass matrix by Adhikary et al.\ \cite{Adhikary:2013bma}, the row-wise elements of $U$ are given in terms of the elements of the $h$ and its eigenvalues $m_i^2$:
\begin{equation}
\label{eq:mod_Ui}
\begin{split}
& U_{1i} = \frac{(h_{22}-m_i^2)h_{13}-h_{12}h_{23}}{N_i}
\\
& U_{2i} = \frac{(h_{11}-m_i^2)h_{23}-h_{12}^*h_{13}}{N_i}
\\
& U_{3i} = \frac{|h_{12}|^2-(h_{11}-m_i^2)(h_{22}-m_i^2)}{N_i},
\end{split}
\end{equation}
where $N_i$ is the normalization constant. Following the PDG convention \cite{Tanabashi:2018oca}, the PMNS matrix $U$ can be parameterized as:
\begin{equation}
\label{eq:PMNS}
U_{PMNS} = 
P_{\phi}
\begin{pmatrix}
c_{12}c_{13} \ & s_{12}c_{13} \ & s_{13}e^{-i\delta}\\
-s_{12}c_{23}-c_{12}s_{23}s_{13}e^{i\delta} \ & c_{12}c_{23}-s_{12}s_{23}s_{13}e^{i\delta} \ & s_{23}c_{13} \\
s_{12}c_{23}-c_{12}c_{23}s_{13}e^{i\delta} \ & -c_{12}s_{23}-s_{12}c_{23}s_{13}e^{i\delta} \ & c_{23}c_{13}
\end{pmatrix}
P_M
\end{equation}
where 
\begin{equation}
\label{eq:mod_Pphi}
P_{\phi} = 
\begin{pmatrix}
e^{i\phi_1} \ & 0 \ & 0\\
0 \ & e^{i\phi_2} \ & 0 \\
0 \ & 0 \ & e^{i\phi_3}
\end{pmatrix}
\end{equation}
contains the unphysical phases that can be rotated away by phase redefinitions of the charged lepton fields, and
\begin{equation}
\label{eq:mod_PM}
P_M = 
\begin{pmatrix}
e^{\frac{i\alpha_M}{2}} \ & 0 \ & 0\\
0 \ & e^{\frac{i\beta_M}{2}} \ & 0 \\
0 \ & 0 \ & 1
\end{pmatrix}
\end{equation}
contains two Majorana phases $\alpha_M$ and $\beta_M$. 
The mixing angles $\theta_{12}$, $\theta_{23}$, and $\theta_{13}$ can be expressed in terms of the elements of $U$ as
\begin{equation}
\label{eq:mod_mix_ang}
 s_{12}^2 = \frac{|U_{12}|^2}{1-|U_{13}|^2}, \ \ \ s_{23}^2 = \frac{|U_{23}|^2}{1-|U_{13}|^2}, \ \ \ s_{13}^2 = |U_{13}|^2.
\end{equation}
The Dirac phase $\delta$ is obtained from the phase redefinition independent quantity $h_{12} h _{23} h_{31}$ through
\begin{equation}
\label{eq:mod_delta}
 \delta = \sin^{-1} \left(\frac{8\, \text{Im}(h_{12} h _{23} h_{31})}{P} \right)
\end{equation}
where P is
\begin{equation}
\label{eq:mod_delta_P}
P = (m^2_2-m^2_1)(m^2_3-m^2_2)(m^2_3-m^2_1)\,\text{sin}2\theta_{12}\,\text{sin}2\theta_{23}\,\text{sin}2\theta_{13}\,\text{cos}\theta_{13}.
\end{equation}
Replacing the elements of $h$ and the $m_i$'s with $H$ and $\eta$ through eq.~\ref{eq:mod_omega}, the mixing angles and the Dirac phase are all ultimately expressed in terms of the modified model parameters. 

To compare with the most recent experiment data on neutrino mixing parameters \cite{deSalas:2017kay}, we fit the modified $A_4$ model to the experimental data by minimizing the following $\chi^2$ function \cite{Rodejohann:2012cf}:
\begin{equation}
\label{eq:chi_sq}
\chi^2 = \sum_i{\frac{(\rho_i-\rho^0_i)^2}{\sigma^2_i}},
\end{equation}
where $\rho^0_i$ is the data of the $i^{\rm th}$ experimental observable, and $\sigma_i$ is the corresponding $1\sigma$ error. $\rho_i$ is the model prediction for the $i^{\rm th}$ observable. The experimental values for all the observables fitted are given in Table \ref{tab:observables}. We fit the six observables to the four free parameters\footnote{The six observables are those listed in Table~\ref{tab:observables}, and the four free parameters are the real and imaginary parts of the complex parameters $H$ and $\eta$ (see eq.\ref{eq:mod_param}).}  of our modified model (i.e., two degrees of freedom). We obtain in the normal hierarchy (NH) $\chi^2_{\rm min} \approx 1.45$, indicating a good description of the NH data. For inverse hierarchy (IH) we obtain $\chi^2_{\rm min} \approx 2.57$, which is somewhat less robust than in the NH case and is consistent with other recent finds that give slight preference in favor of the normal hierarchy over the inverse hierarchy~\cite{deSalas:2017kay}. 

\begin{figure}[tbp]
\centering 
\includegraphics[width=\linewidth]{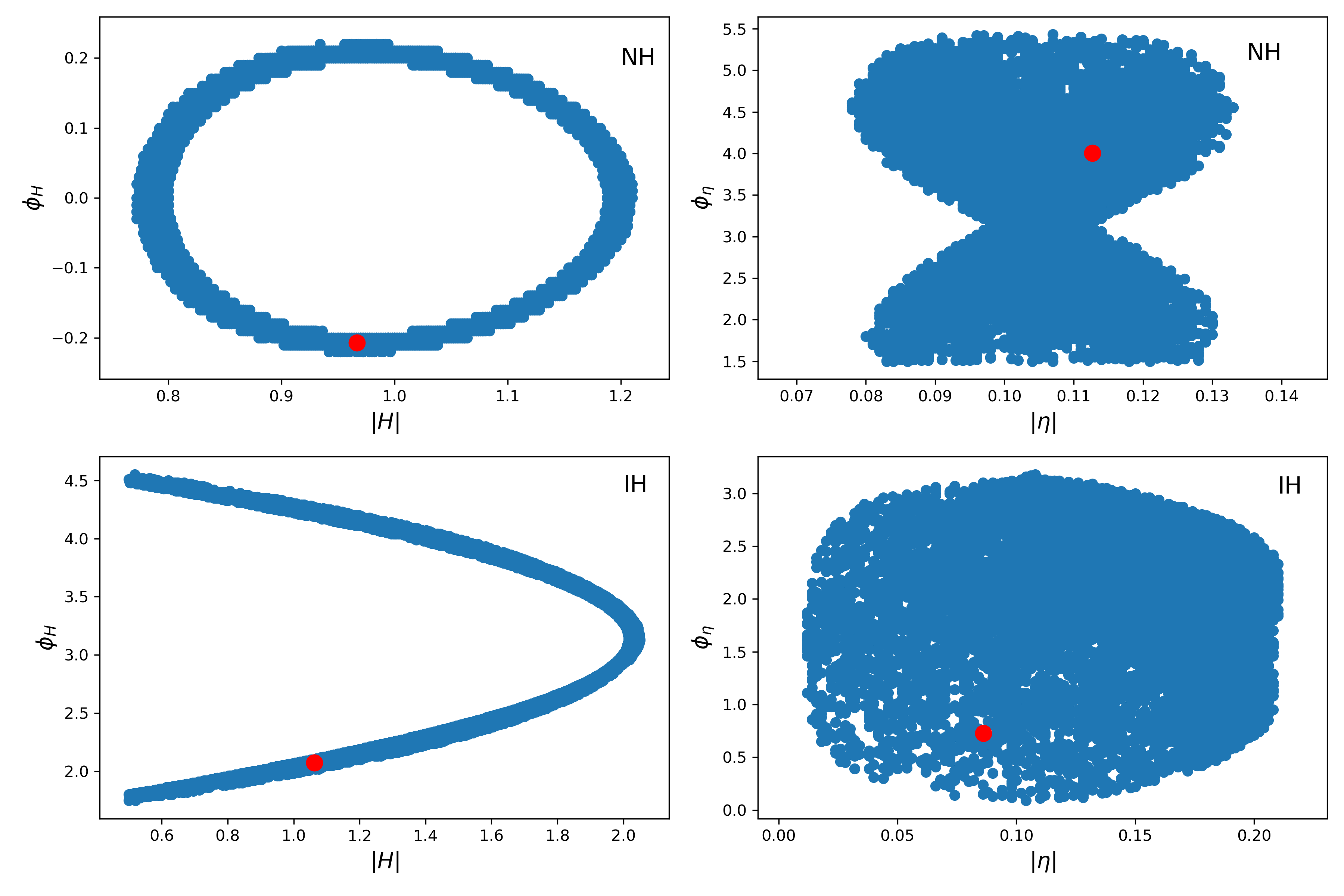}
\caption{\label{fig:pred_range} The allowed region of the model parameters $|H|$, $\phi_{H}$, $|\eta|$, $\phi_{\eta}$ at 3$\sigma$ C.L. in both NH and IH cases. Here the C.L. ranges are defined as contours in $\Delta^2$ with respect to $\chi^2_{\rm min}$. The best fit values with $\chi^2 = \chi^2_{\rm min}$ are indicated by the red dots. $\chi^2_{\rm min} \approx 1.45$ in NH and $\chi^2_{\rm min} \approx 2.57$ in IH. }
\end{figure}

Fig.~\ref{fig:pred_range} shows the allowed regions of model parameters $|H|$, $\phi_{H}$, $|\eta|$, $\phi_{\eta}$ at $1\sigma$, $2\sigma$, and $3\sigma$ C.L., defined as contours in $\Delta \chi^2$ with respect to $\chi^2_{\rm min}$. Here $|z|$ and $\phi_z$ are the amplitude and phase respectively of the complex parameter $z$. The best fit values for  $|H|$, $\phi_{H}$, $|\eta|$, and $\phi_{\eta}$ are $(0.967, 1.934\pi, 0.113, 1.275\pi)$ in NH and $(1.063 , 0.660\pi, 0.086, 0.212\pi)$ in IH.

\begin{figure}[tbp]
\centering 
\includegraphics[width=\linewidth]{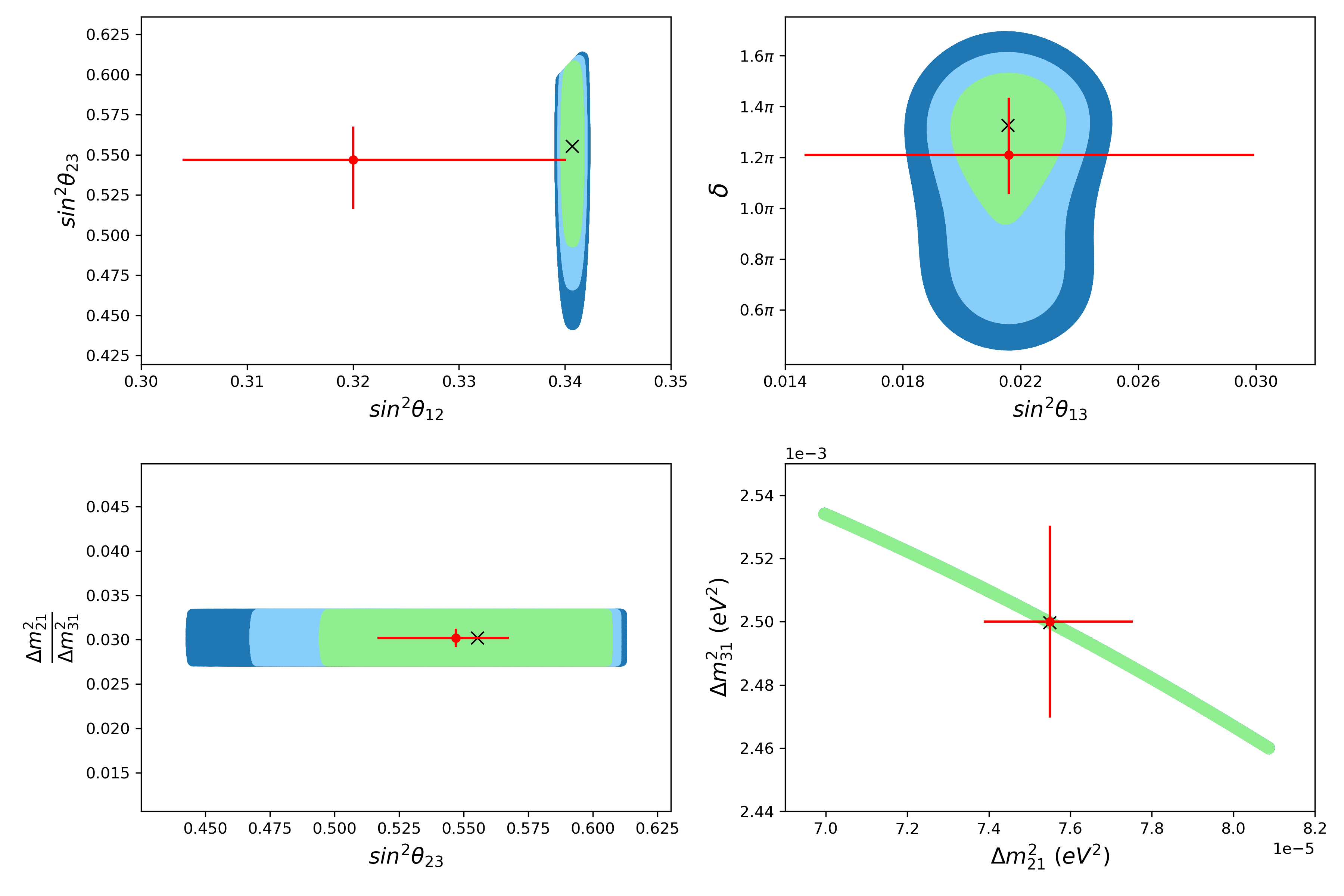}
\caption{\label{fig:pred_nh} The allowed region of the oscillation parameters predicted by the model at 1$\sigma$ (green), 2$\sigma$ (light blue), 3$\sigma$ (dark blue) C.L. in the normal hierarchy case. The best fit of the model is indicted by the black x. For comparison, the experimental best fit data and 1$\sigma$ range are indicted by red dot and vertical/horizontal bars. The best fit gives $\chi^2_{\rm min} \approx$ 1.45, indicating good agreement with current data with nonzero $\theta_{13}$ and Dirac phase $\delta$, improving from original tri-bimaximal mixing model. }
\end{figure}

\begin{figure}[tbp]
\centering 
\includegraphics[width=\linewidth]{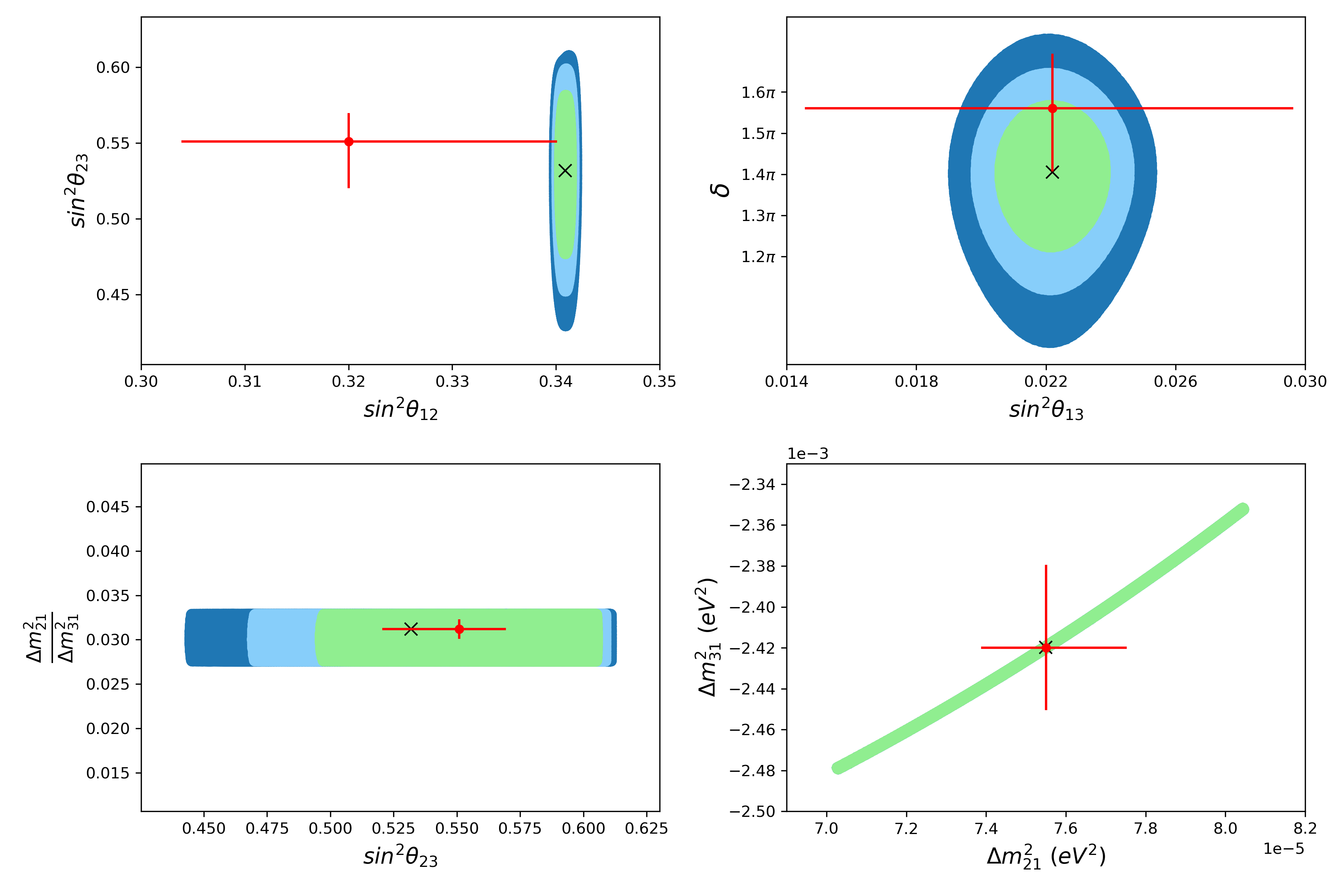}
\caption{\label{fig:pred_ih} The allowed region of the oscillation parameters predicted by the model at 1$\sigma$ (green), 2$\sigma$ (light blue), 3$\sigma$ (dark blue) C.L. in the inverse hierarchy case. The best fit of the model is indicted by the black x. For comparison, the experimental best fit data and 1$\sigma$ range are indicted by red dot and vertical/horizontal bars. The best fit gives $\chi^2_{\rm min} \approx$ 2.57, a better description than the original tri-bimaximal model , but less compatible with experimental data than in the NH case. }
\end{figure}

In Figs.~\ref{fig:pred_nh} and \ref{fig:pred_ih}, the predictions for the oscillation parameters and neutrino mass squared differences are presented for both NH and IH cases. In NH, the best fit values of $\sin\theta_{23}$ and $\sin\theta_{13}$ are $(0.555, 0.0216)$, in good agreement with experimental values of $(0.547^{+0.020}_{-0.030}, 0.0216^{+0.0083}_{-0.0069})$. The best fit value of $\delta$ is $1.32\pi$, slightly above the experimental value at $1.21\pi$, but well within the $1\sigma$ range. The model prediction for $\sin\theta_{12}$ is narrowly centered around $0.340$, at the upper edge of the $1\sigma$ range of the experimental value centered at $0.320$. This is a sharp prediction of the model that can be tested as experimental uncertainty on the mixing angles narrows. The mass squared differences $\Delta m_{21}^2$ and $\Delta m_{31}^2$ are also fitted and show excellent agreement with the data. The predictions in the IH case are similar, with the predicted value of $\sin\theta_{23}$ and $\delta$ lying on the lower edge of the $1\sigma$ range of the global fit to experimental values. Thus, the modified Altarelli-Feruglio model with the additional contraction term in eq.~\ref{eq:mod_term} produces deviation from the tri-bimaxial matrix that is in good agreement with recent experimental data on neutrino oscillation parameters with nonzero $\theta_{13}$ and Dirac phase $\delta$. 

\begin{table}[tbp]
\centering
\begin{tabular}{cccc}
\hline Parameter                                      & Best fit $\pm 1\sigma$    & $2\sigma$ range & $3\sigma$ range \\ \hline
$\Delta m^2_{21} [10^{-5} eV^2]$        & $7.55^{+0.20}_{-0.16}$    & 7.20-7.94       & 7.05-8.14       \\[0.1cm]
$|\Delta m^2_{31}| [10^{-3} eV^2]$ (NH) & $2.50\pm0.03$             & 2.44-2.57       & 2.41-2.60       \\[0.1cm]
$|\Delta m^2_{31}| [10^{-3} eV^2]$ (IH) & $2.42^{+0.03}_{-0.04}$    & 2.34-2.47       & 2.31-2.51       \\[0.1cm]
$\sin^2 \theta_{12} / 10^{-1}$                  & $3.20^{+0.20}_{-0.16}$    & 2.89-3.59       & 2.73-3.79       \\[0.1cm]
$\sin^2 \theta_{23} / 10^{-1}$ (NH)             & $5.47^{+0.20}_{-0.30}$    & 4.67-5.83       & 4.45-5.99       \\[0.1cm]
$\sin^2 \theta_{23} / 10^{-1}$ (IH)             & $5.51^{+0.18}_{-0.30}$    & 4.91-5.84       & 4.53-5.98       \\[0.1cm]
$\sin^2 \theta_{13} / 10^{-2}$ (NH)             & $2.160^{+0.083}_{-0.069}$ & 2.03-2.34       & 1.96-2.41       \\[0.1cm]
$\sin^2 \theta_{13} / 10^{-2}$ (IH)             & $2.220^{+0.074}_{-0.076}$ & 2.07-2.36       & 1.99-2.44       \\[0.1cm]
$\delta / \pi$ (NH)                            & $1.21^{+0.21}_{-0.15}$    & 1.01-1.75       & 0.87-1.94       \\[0.1cm]
$\delta / \pi$ (IH)                            & $1.56^{+0.13}_{-0.15}$    & 1.27-1.82       & 1.12-1.94          \\ \hline
\end{tabular}
\caption{Neutrino oscillation parameters determined from global analysis of experimental data \cite{deSalas:2017kay}.}
\label{tab:observables}
\end{table}

\subsection{Prediction for $J$ and Effective Majorana Mass}

\begin{figure}[tbp]
\centering 
\includegraphics[width=\linewidth]{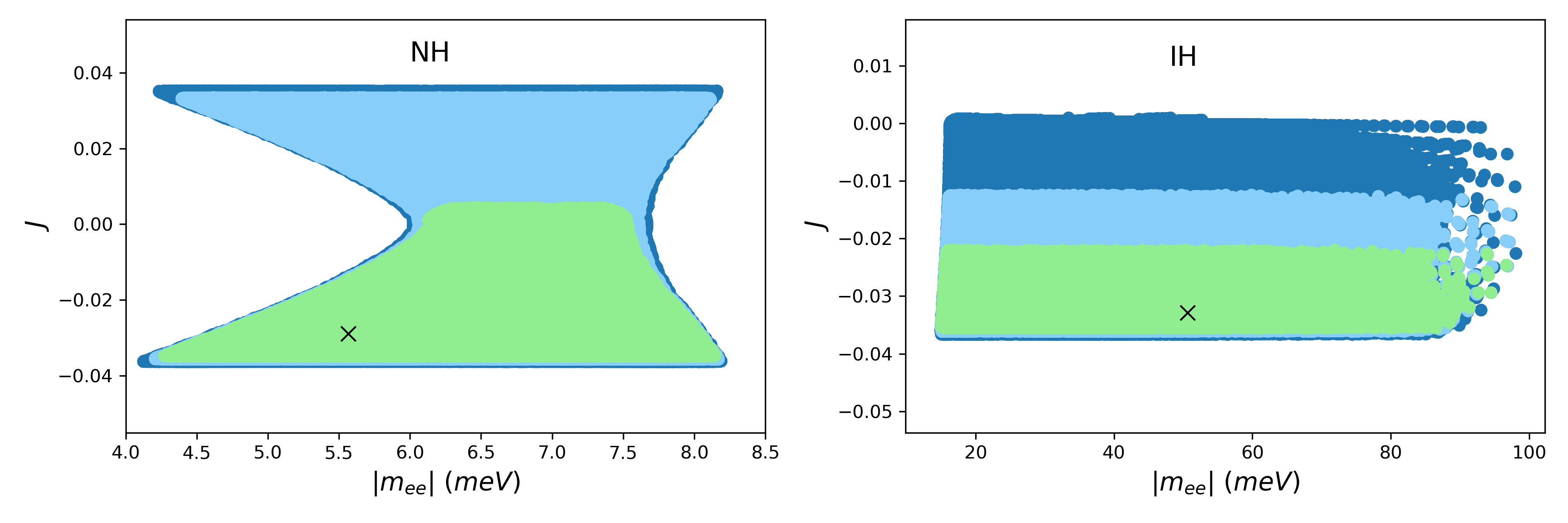}
\caption{\label{fig:cp_param} The allowed region of the Jarlskog invariant ($J$) and effective Majorana neutrino mass ($| m_{ee} |$), predicted by the model at 1$\sigma$ (green), 2$\sigma$ (light blue), 3$\sigma$ (dark blue) C.L. for both NH and IH. The best fit of the model is indicted by the black x. The current upper bound of $| m_{ee} | = 0.16\, {\rm eV}$ is indicted by the red dotted vertical line. The predicted $|m_{ee}|$ range in IH can be tested with near-future ${0\nu\beta\beta}$ experiments, but the NH predictions are too small for planned detector sensitivities. }
\end{figure}

In Fig.~\ref{fig:cp_param} we present the additional predictions of the modified model on CP-violation Jarlskog parameter $J$, and the effective Majorana mass $|m_{ee}|$, characterizing $0\nu\beta\beta$ decay. The Jarlskog invariant is given by the phase redefinition invariant quantity
\begin{equation}
\label{eq:jarls}
J = \text{Im}\{U_{11}U_{22}U_{12}^{*}U_{21}^{*}\} = s_{12}c_{13}^2 s_{12}c_{12}s_{23}c_{23}\sin\delta,
\end{equation}
and the effective Majorana mass is given by
\begin{equation}
\label{eq:m_ee}
|m_{ee}| = |\sum_i U^2_{1i} m_i| = |c^2_{12}c^2_{13} m_1 e^{i\alpha_M} + s^2_{12}c^2_{13} m_2 e^{i\beta_M} + s^2_{13}m_3 e^{-i2\delta}| = |(m_{\nu})_{11}|.
\end{equation}
$J$ is completely specified by the mixing angles and the Dirac phase discussed in the previous section. The effective Majorana mass contains the Majorana phases $\alpha_M$ and $\beta_M$. Our obtained $3\sigma$ range of values for $|m_{ee}|$ is 4.26 meV $\sim$ 8.13 meV in NH, compared to $|m_{ee}| \approx$ 7 meV in the original Altarelli-Feruglio model; and 15.0 meV $\sim$ 95.6 meV in IH, compared to $|m_{ee}| \geq$ 17 meV. The effective Majorana mass has the upper bound of $|m_{ee}| \leq$ 160 meV, corresponding to $T^{0\nu\beta\beta}_{1/2} ({}^{136}{\rm Xe}) \geq 1.1 \times 10^{26}$ yr at 90\% C.L, which follows from the data of the KamLAND-Zen experiment \cite{KamLAND-Zen:2016pfg}. For IH, the model predicted $|m_{ee}|$ range is within the sensitivity of near future $0\nu\beta\beta$ decay experiments. For example, the bolometric CUORE experiment, using ${}^{130}{\rm Te}$, has a sensitivity of $|m_{ee}| \leq$ 50 meV, which corresponds to $T^{0\nu\beta\beta}_{1/2} ({}^{130}{\rm Te}) \geq 10^{26}$ yr \cite{Alessandria:2011rc}. There are also planned ton-scale next-to-next generation $0\nu\beta\beta$ experiments using ${}^{136}{\rm Xe}$ \cite{KamLANDZen:2012aa, Albert:2014fya} and ${}^{76}{\rm Ge}$ \cite{Abt:2004yk, Guiseppe:2011me} that can reach a sensitivity of $|m_{ee}| \sim$ 12 - 30 meV, corresponding to $T^{0\nu\beta\beta}_{1/2} \geq 10^{27}$ yr \cite{CarcamoHernandez:2017kra}. However, this sensitivity is still outside the range of model predicted $|m_{ee}|$ values in NH, which is too small to be tested in the immediate next-generation $0\nu\beta\beta$ experiments. 

\section{Conclusion}

In summary, we proposed a simple modification to the Altarelli-Feruglio $A_4$ flavor symmetry model that retains the minimal structure of the original model, while predicting oscillation parameters in much better agreements with current experimental data. The best model-data agreement is found in the normal hierarchy case, with $\chi^2_{\rm min} \approx 1.45$, consistent with recent data analyses in favor of the normal hierarchy over the inverse hierarchy \cite{deSalas:2017kay}. The prediction of $\sin\theta_{12}$ narrowly centered around 0.34 can be tested in the future as experimental uncertainty on the mixing angles narrows in the next round of experiments. The prediction of the modified model on CP-violating parameters and effective Majorana mass are also investigated. The predicted range of $|m_{ee}|$ in the IH case are within the sensitivity of planned detectors, whereas the NH predicted values are still too small to be tested in the near future. In the process of studying the modified model, we systematically worked out the $A_4$ field contraction projection matrices in the different cases of Dirac ($\chi^{\dagger} \tau\psi$) vs. Majorana ($\chi^{T} \tau\psi$) fields and $S$- vs. $T$-diagonal bases. All cases are listed in the appendix for future reference. 

\appendix
\section{$A_4$ Group Contractions}

\subsection{Basics of the $A_4$ Group}

$A_4$ is the group of even permutation of 4 elements, or geometrically, the symmetry group of a tetrahedron. It has $(4 !)/2=12$ elements, which can all be generated by two basic elements $S$ and $T$ given in permutation form by $S=(4321)$ and $T=(2314)$. $S$ and $T$ satisfy the properties
\begin{equation}
\label{eq:a4_gen}
S^2 = T^3 = (ST)^3 = 1
\end{equation}
which specify the presentation of the group

By looking at the equivalence classes of the $A_4$ group and the characters of elements in each class, one can derive that $A_4$ has four inequivalent irreducible representations: three singlet representations $1$, $1'$, and $1''$ and one triplet representation $3$. In the singlet representations, the generator $S$ and $T$ are given by
\begin{equation}
\label{eq:a4_rep}
\begin{split}
&1:\ \ \ S = 1\ \ T = 1
\\
&1':\ \ S = 1\ \ T = e^{i2\pi/3} \equiv \omega
\\
&1'':\ S = 1\ \ T = e^{i4\pi/3} \equiv \omega^2.
\end{split}
\end{equation}
In the triplet representation, in the basis where $S$ is diagonal, the generators are given by
\begin{equation}
\label{eq:s_diag}
S' = 
\begin{pmatrix}
1 \ & 0 & 0\\
0 \ & -1 & 0\\
0 \ & 0 & -1
\end{pmatrix}, \ \
T' = 
\begin{pmatrix}
0 \ & 1 \  & 0\\
0 \ & 0 \ & 1\\
1 \ & 0 \ & 0
\end{pmatrix}
\end{equation}
In the $T$ diagonal basis the generators of the triplet representation are
\begin{equation}
\label{eq:t_diag}
S = VS'V^{\dagger} = 
\begin{pmatrix}
-1 \ & 2 \  & 2\\
2 \ & -1 \ & 2\\
2 \ & 2 \ & -1
\end{pmatrix}, \ \
T = VT'V^{\dagger} = \frac{1}{3}
\begin{pmatrix}
1 \ & 0 & 0\\
0 \ & \omega & 0\\
0 \ & 0 & \omega^2
\end{pmatrix}
\end{equation}
with the unitary transformation matrix $V$ given by
\begin{equation}
\label{eq:v_mat}
V = \frac{1}{\sqrt{3}}
\begin{pmatrix}
1 \ & 1 & 1\\
0 \ & \omega^2 & \omega\\
0 \ & \omega & \omega^2
\end{pmatrix}.
\end{equation}
A comprehensive review of the properties of the $A_4$ group and other non-abelian discrete symmetries can be found in ref.~\cite{Ishimori:2010au}. 

\subsection{Tensor Product and Field Contraction}
The tensor product rule for $A_4$ representations are given by
\begin{equation}
\label{eq:tensor_prod}
\begin{split}
&1 \times 1 = 1; \ \ \ \ \ 1 \times 1' = 1'; \ \ \ 1 \times 1'' = 1'';
\\
&1' \times 1' = 1''; \ \ 1' \times 1''= 1; \ \ 1'' \times 1'' = 1';
\\
&1/1/1'' \times 3 = 3 \times 1/1'/1'' = 3;
\\
& 3 \times 3 = 1 + 1' + 1'' + 3_1 + 3_2.
\end{split}
\end{equation}
In the context of $A_4$ flavor symmetry models, the tensor product rules specify the contraction rules of two particle fields $\chi$ and $\psi$ under $A_4$: 
\begin{equation}
\label{eq:field_contrac}
\begin{split}
& \chi^{\dagger} \tau\psi = \eta \ \ \ \text{(Dirac Fields)} \\
& \chi^{T} \tau\psi = \eta \ \ \text{(Majorana Fields)}
\end{split}
\end{equation}
where $\eta$ is the result of the contraction and is a particular $A_4$  representation specified by the tensor product rules. $\tau$ is the projection matrix that determines the components of $\eta$ in terms of the components of $\chi$ and $\psi$. Thus, it is important for $A_4$ flavor symmetry models to know the form of the projection matrix $\tau$ in each tensor product combination, which can be different in $S$-diagonal or $T$-diagonal basis, and for Dirac vs.\ Majorana fields. The forms of the projection matrices for each of these different scenarios are summarized in the next section.

\subsection{Form of Projection Matrix}
\label{sec:proj_mat}

Here we summarize the form of the projection matrix $\tau$ for Dirac ($\chi^{\dagger} \tau\psi$) vs.\ Majorana ($\chi^{T} \tau\psi$) fields, in $S$-diagonal or $T$-diagonal basis.
 
\subsubsection{Only Singlets}

The contractions involving only singlet representations $1$, $1'$, and $1''$ are trivial and have $\tau = 1$ for all cases of Dirac vs.\ Majorana and $S$-diagonal vs.\  $T$-diagonal.  

\subsubsection{Singlet and Triplet}

For the trivial contractions $1 \times 3 = 3$ and $3 \times 1 = 3$ where a singlet $1$ is involved, $\tau = I_{3}$, the $3 \times 3$ identity matrix in all cases. 

For $1' \times 3 = 3$, we have
\begin{equation}
\label{eq:tau_1p3}
\begin{tabular}{cccc}
\multicolumn{2}{c}{Dirac ($\chi^{\dagger} \tau\psi$)} & \multicolumn{2}{c}{Majorana ($\chi^{T} \tau\psi$)} \\
$\tau_S$ =
$\begin{pmatrix}
1 \ & 0 & 0\\
0 \ & \omega^2 & 0 \\
0 \ & 0 & \omega
\end{pmatrix}$
&    
$\tau_T$ =
$\begin{pmatrix}
0 \ & 1 & 0 \\
0 \ & 0 & 1 \\
1 \ & 0 & 0
\end{pmatrix};$
&     
$\tau_S$ =
$\begin{pmatrix}
1 \ & 0 & 0\\
0 \ & \omega & 0 \\
0 \ & 0 & \omega^2
\end{pmatrix}$     
&    
$\tau_T$ =
$\begin{pmatrix}
0 \ & 0 & 1 \\
0 \ & 1 & 0 \\
1 \ & 0 & 0
\end{pmatrix}$   
\end{tabular}
\end{equation}
where $\tau_S$ and $\tau_T$ denote the projection matrix in $S$ and $T$-diagonal bases respectively.

$ 3 \times 1' = 3$:
\begin{equation}
\label{eq:tau_31p}
\begin{tabular}{cccc}
\multicolumn{2}{c}{Dirac ($\chi^{\dagger} \tau\psi$)} & \multicolumn{2}{c}{Majorana ($\chi^{T} \tau\psi$)} \\
$\tau_S$ =
$\begin{pmatrix}
1 \ & 0 & 0\\
0 \ & \omega & 0 \\
0 \ & 0 & \omega^2
\end{pmatrix}$
&    
$\tau_T$ =
$\begin{pmatrix}
0 \ & 0 & 1 \\
1 \ & 0 & 0 \\
0 \ & 1 & 0
\end{pmatrix};$
&     
$\tau_S$ =
$\begin{pmatrix}
1 \ & 0 & 0\\
0 \ & \omega & 0 \\
0 \ & 0 & \omega^2
\end{pmatrix}$     
&    
$\tau_T$ =
$\begin{pmatrix}
0 \ & 0 & 1 \\
0 \ & 1 & 0 \\
1 \ & 0 & 0
\end{pmatrix}$   
\end{tabular}
\end{equation}

$ 1'' \times 3 = 3$:
\begin{equation}
\label{eq:tau_1pp3}
\begin{tabular}{cccc}
\multicolumn{2}{c}{Dirac ($\chi^{\dagger} \tau\psi$)} & \multicolumn{2}{c}{Majorana ($\chi^{T} \tau\psi$)} \\
$\tau_S$ =
$\begin{pmatrix}
1 \ & 0 & 0\\
0 \ & \omega & 0 \\
0 \ & 0 & \omega^2
\end{pmatrix}$
&    
$\tau_T$ =
$\begin{pmatrix}
0 \ & 0 & 1 \\
1 \ & 0 & 0 \\
0 \ & 1 & 0
\end{pmatrix};$
&     
$\tau_S$ =
$\begin{pmatrix}
1 \ & 0 & 0\\
0 \ & \omega^2 & 0 \\
0 \ & 0 & \omega
\end{pmatrix}$     
&    
$\tau_T$ =
$\begin{pmatrix}
0 \ & 1 & 0 \\
1 \ & 0 & 0 \\
0 \ & 0 & 1
\end{pmatrix}$   
\end{tabular}
\end{equation}

$ 3 \times 1'' = 3$:
\begin{equation}
\label{eq:tau_31pp}
\begin{tabular}{cccc}
\multicolumn{2}{c}{Dirac ($\chi^{\dagger} \tau\psi$)} & \multicolumn{2}{c}{Majorana ($\chi^{T} \tau\psi$)} \\
$\tau_S$ =
$\begin{pmatrix}
1 \ & 0 & 0\\
0 \ & \omega^2 & 0 \\
0 \ & 0 & \omega
\end{pmatrix}$
&    
$\tau_T$ =
$\begin{pmatrix}
0 \ & 1 & 0 \\
0 \ & 0 & 1 \\
1 \ & 0 & 0
\end{pmatrix};$
&     
$\tau_S$ =
$\begin{pmatrix}
1 \ & 0 & 0\\
0 \ & \omega^2 & 0 \\
0 \ & 0 & \omega
\end{pmatrix}$     
&    
$\tau_T$ =
$\begin{pmatrix}
0 \ & 1 & 0 \\
1 \ & 0 & 0 \\
0 \ & 0 & 1
\end{pmatrix}$   
\end{tabular}
\end{equation}

\subsubsection{Only Triplets}
Here we list the form of the project matrices for all triplet contractions. In the cases where two triplets contract to another triplet, there are two linearly independent contractions with different $\tau^{\rho}_{\mu\nu}$. 

$3 \times 3 = 1$:
\begin{equation}
\label{eq:tau_331}
\begin{tabular}{cccc}
\multicolumn{2}{c}{Dirac ($\chi^{\dagger} \tau\psi$)} & \multicolumn{2}{c}{Majorana ($\chi^{T} \tau\psi$)} \\
\multicolumn{2}{c}{
$\tau_S$ = $\tau_T$ = 
$\begin{pmatrix}
1 \ & 0 & 0 \\
0 \ & 1 & 0 \\
0 \ & 0 & 1
\end{pmatrix};$
}
&     
$\tau_S$ =
$\begin{pmatrix}
1 \ & 0 & 0 \\
0 \ & 1 & 0 \\
0 \ & 0 & 1
\end{pmatrix}$     
&    
$\tau_T$ =
$\begin{pmatrix}
1 \ & 0 & 0 \\
0 \ & 0 & 1 \\
0 \ & 1 & 0
\end{pmatrix}$   
\end{tabular}
\end{equation}

$3 \times 3 = 1'$:
\begin{equation}
\label{eq:tau_331p}
\begin{tabular}{cccc}
\multicolumn{2}{c}{Dirac ($\chi^{\dagger} \tau\psi$)} & \multicolumn{2}{c}{Majorana ($\chi^{T} \tau\psi$)} \\
$\tau_S$ =
$\begin{pmatrix}
1 \ & 0 & 0\\
0 \ & \omega^2 & 0 \\
0 \ & 0 & \omega
\end{pmatrix}$
&    
$\tau_T$ =
$\begin{pmatrix}
0 \ & 1 & 0 \\
0 \ & 0 & 1 \\
1 \ & 0 & 0
\end{pmatrix};$
&     
$\tau_S$ =
$\begin{pmatrix}
1 \ & 0 & 0\\
0 \ & \omega^2 & 0 \\
0 \ & 0 & \omega
\end{pmatrix}$     
&    
$\tau_T$ =
$\begin{pmatrix}
0 \ & 1 & 0 \\
1 \ & 0 & 0 \\
0 \ & 0 & 1
\end{pmatrix}$   
\end{tabular}
\end{equation}

$3 \times 3 = 1''$:
\begin{equation}
\label{eq:tau_331pp}
\begin{tabular}{cccc}
\multicolumn{2}{c}{Dirac ($\chi^{\dagger} \tau\psi$)} & \multicolumn{2}{c}{Majorana ($\chi^{T} \tau\psi$)} \\
$\tau_S$ =
$\begin{pmatrix}
1 \ & 0 & 0\\
0 \ & \omega & 0 \\
0 \ & 0 & \omega^2
\end{pmatrix}$
&    
$\tau_T$ =
$\begin{pmatrix}
0 \ & 0 & 1 \\
1 \ & 0 & 0 \\
0 \ & 1 & 0
\end{pmatrix};$
&     
$\tau_S$ =
$\begin{pmatrix}
1 \ & 0 & 0\\
0 \ & \omega & 0 \\
0 \ & 0 & \omega^2
\end{pmatrix}$     
&    
$\tau_T$ =
$\begin{pmatrix}
0 \ & 0 & 1 \\
0 \ & 1 & 0 \\
1 \ & 0 & 0
\end{pmatrix}$   
\end{tabular}
\end{equation}

$3 \times 3 = 3_1$:
\begin{equation}
\label{eq:tau_3331}
\begin{tabular}{clcl}
\multicolumn{2}{c}{Dirac ($\chi^{\dagger} \tau\psi$)} \\[0.1cm]
\multicolumn{2}{c}{
${\tau_S}^{\rho}_{\mu\nu}$ = 
$\left[
\begin{pmatrix}
0 & 0 & 0\\
0 & 0 & 1 \\
0 & 0 & 0
\end{pmatrix},
\begin{pmatrix}
0 & 0 & 0\\
0 & 0 & 0 \\
1 & 0 & 0
\end{pmatrix},
\begin{pmatrix}
0 & 1 & 0\\
0 & 0 & 0 \\
0 & 0 & 0
\end{pmatrix}
\right]$
} \\[1cm]    
\multicolumn{2}{c}{
${\tau_T}^{\rho}_{\mu\nu}$ = 
$\left[
\begin{pmatrix}
2 & 0 & 0 \\
0 & -1 & 0 \\
0 & 0 & -1
\end{pmatrix},
\begin{pmatrix}
0 & -1 & 0\\
0 & 0 & 2 \\
-1 & 0 & 0
\end{pmatrix},
\begin{pmatrix}
0 & 0 & -1\\
-1 & 0 & 0 \\
0 & 2 & 0
\end{pmatrix}
\right]$
} \\[1cm]
\multicolumn{2}{c}{Majorana ($\chi^{T} \tau\psi$)} \\[0.1cm]    
\multicolumn{2}{c}{
${\tau_S}^{\rho}_{\mu\nu}$ = 
$\left[
\begin{pmatrix}
0 & 0 & 0\\
0 & 0 & 1 \\
0 & 0 & 0
\end{pmatrix},
\begin{pmatrix}
0 & 0 & 0\\
0 & 0 & 0 \\
1 & 0 & 0
\end{pmatrix},
\begin{pmatrix}
0 & 1 & 0\\
0 & 0 & 0 \\
0 & 0 & 0
\end{pmatrix}
\right]$
} \\[1cm]
\multicolumn{2}{c}{
${\tau_T}^{\rho}_{\mu\nu}$ = 
$\left[
\begin{pmatrix}
2 & 0 & 0\\
0 & 0 & -1 \\
0 & -1 & 0
\end{pmatrix},
\begin{pmatrix}
0 & -1 & 0\\
-1 & 0 & 0 \\
0 & 0 & 2
\end{pmatrix},
\begin{pmatrix}
0 & 0 & -1\\
0 & 2 & 0 \\
-1 & 0 & 0
\end{pmatrix}
\right]$
}
\end{tabular}
\end{equation}

$3 \times 3 = 3_2$:
\begin{equation}
\label{eq:tau_3332}
\begin{tabular}{clcl}
\multicolumn{2}{c}{Dirac ($\chi^{\dagger} \tau\psi$)} \\[0.1cm]
\multicolumn{2}{c}{
${\tau_S}^{\rho}_{\mu\nu}$ = 
$\left[
\begin{pmatrix}
0 & 0 & 0\\
0 & 0 & 0 \\
0 & 1 & 0
\end{pmatrix},
\begin{pmatrix}
0 & 0 & 1\\
0 & 0 & 0 \\
0 & 0 & 0
\end{pmatrix},
\begin{pmatrix}
0 & 0 & 0\\
1 & 0 & 0 \\
0 & 0 & 0
\end{pmatrix}
\right]$
} \\[1cm] 
\multicolumn{2}{c}{
${\tau_T}^{\rho}_{\mu\nu}$ = 
$\left[
\begin{pmatrix}
0 & 0 & 0 \\
0 & 1 & 0 \\
0 & 0 & -1
\end{pmatrix},
\begin{pmatrix}
0 & -1 & 0\\
0 & 0 & 0 \\
1 & 0 & 0
\end{pmatrix},
\begin{pmatrix}
0 & 0 & 1\\
-1 & 0 & 0 \\
0 & 0 & 0
\end{pmatrix}
\right]$
} \\[1cm]
\multicolumn{2}{c}{Majorana ($\chi^{T} \tau\psi$)} \\[0.1cm]
\multicolumn{2}{c}{
${\tau_S}^{\rho}_{\mu\nu}$ = 
$\left[
\begin{pmatrix}
0 & 0 & 0\\
0 & 0 & 0 \\
0 & 1 & 0
\end{pmatrix},
\begin{pmatrix}
0 & 0 & 1 \\
0 & 0 & 0 \\
0 & 0 & 0
\end{pmatrix},
\begin{pmatrix}
0 & 0 & 0 \\
1 & 0 & 0 \\
0 & 0 & 0
\end{pmatrix}
\right]$
} \\[1cm]
\multicolumn{2}{c}{
${\tau_T}^{\rho}_{\mu\nu}$ = 
$\left[
\begin{pmatrix}
0 & 0 & 0\\
0 & 0 & 1 \\
0 & -1 & 0
\end{pmatrix},
\begin{pmatrix}
0 & 1 & 0\\
-1 & 0 & 0 \\
0 & 0 & 0
\end{pmatrix},
\begin{pmatrix}
0 & 0 & -1\\
0 & 0 & 0 \\
1 & 0 & 0
\end{pmatrix}
\right]$
}
\end{tabular}
\end{equation}

\acknowledgments

We wish to thank B. Shakya for discussions. JDW is supported in part by DOE de-sc0007859. ML and JDW are also supported in part by the LCTP.


\bibliographystyle{JHEP}
\bibliography{reference}

\providecommand{\href}[2]{#2}\begingroup\raggedright\begin{thebibliography}{10}

\bibitem{Altarelli:2010gt}
G.~Altarelli and F.~Feruglio, {\it {Discrete Flavor Symmetries and Models of
  Neutrino Mixing}},  {\em Rev. Mod. Phys.} {\bf 82} (2010) 2701--2729,
  [\href{http://arxiv.org/abs/1002.0211}{{\tt arXiv:1002.0211}}].

\bibitem{Abe:2011fz}
{\bf Double Chooz} Collaboration, Y.~Abe et~al., {\it {Indication of Reactor
  $\bar{\nu}_e$ Disappearance in the Double Chooz Experiment}},  {\em Phys.
  Rev. Lett.} {\bf 108} (2012) 131801,
  [\href{http://arxiv.org/abs/1112.6353}{{\tt arXiv:1112.6353}}].

\bibitem{Ahn:2012nd}
{\bf RENO} Collaboration, J.~K. Ahn et~al., {\it {Observation of Reactor
  Electron Antineutrino Disappearance in the RENO Experiment}},  {\em Phys.
  Rev. Lett.} {\bf 108} (2012) 191802,
  [\href{http://arxiv.org/abs/1204.0626}{{\tt arXiv:1204.0626}}].

\bibitem{Ling:2013fta}
{\bf Daya Bay} Collaboration, J.~Ling, {\it {Observation of
  electron-antineutrino disappearance at Daya Bay}},  {\em AIP Conf. Proc.}
  {\bf 1560} (2013), no.~1 206--210.

\bibitem{Altarelli:2009kr}
G.~Altarelli and D.~Meloni, {\it {A Simplest A4 Model for Tri-Bimaximal
  Neutrino Mixing}},  {\em J. Phys.} {\bf G36} (2009) 085005,
  [\href{http://arxiv.org/abs/0905.0620}{{\tt arXiv:0905.0620}}].

\bibitem{Hall:2013yha}
L.~J. Hall and G.~G. Ross, {\it {Discrete Symmetries and Neutrino Mass
  Perturbations for $\theta_{13}$}},  {\em JHEP} {\bf 11} (2013) 091,
  [\href{http://arxiv.org/abs/1303.6962}{{\tt arXiv:1303.6962}}].

\bibitem{Felipe:2013vwa}
R.~Gonzalez~Felipe, H.~Serodio, and J.~P. Silva, {\it {Neutrino masses and
  mixing in A4 models with three Higgs doublets}},  {\em Phys. Rev.} {\bf D88}
  (2013), no.~1 015015, [\href{http://arxiv.org/abs/1304.3468}{{\tt
  arXiv:1304.3468}}].

\bibitem{Hollik:2017get}
W.~G. Hollik and U.~J. Saldana-Salazar, {\it {Texture zeros and hierarchical
  masses from flavour (mis)alignment}},  {\em Nucl. Phys. B} {\bf 928} (2018)
  535--554, [\href{http://arxiv.org/abs/1712.05387}{{\tt arXiv:1712.05387}}].

\bibitem{Holthausen:2012wt}
M.~Holthausen, K.~S. Lim, and M.~Lindner, {\it {Lepton Mixing Patterns from a
  Scan of Finite Discrete Groups}},  {\em Phys. Lett.} {\bf B721} (2013)
  61--67, [\href{http://arxiv.org/abs/1212.2411}{{\tt arXiv:1212.2411}}].

\bibitem{Ahn:2013ema}
Y.~Ahn, {\it Leptons and quarks from a discrete flavor symmetry},  {\em
  Phys.Rev.D} {\bf 87} (2013), no.~11 113011,
  [\href{http://arxiv.org/abs/1303.4863}{{\tt arXiv:1303.4863}}].

\bibitem{Perez:2019aqq}
M.~J. Pérez, M.~H. Rahat, P.~Ramond, A.~J. Stuart, and B.~Xu, {\it {Stitching
  an asymmetric texture with $\mathcal{T}_{13} \times \mathcal{Z}_5$ family
  symmetry}},  {\em Phys. Rev. D} {\bf 100} (2019), no.~7 075008,
  [\href{http://arxiv.org/abs/1907.10698}{{\tt arXiv:1907.10698}}].

\bibitem{Rahat:2018sgs}
M.~H. Rahat, P.~Ramond, and B.~Xu, {\it {Asymmetric tribimaximal texture}},
  {\em Phys. Rev. D} {\bf 98} (2018), no.~5 055030,
  [\href{http://arxiv.org/abs/1805.10684}{{\tt arXiv:1805.10684}}].

\bibitem{Adhikary:2013bma}
B.~Adhikary, M.~Chakraborty, and A.~Ghosal, {\it {Masses, mixing angles and
  phases of general Majorana neutrino mass matrix}},  {\em JHEP} {\bf 10}
  (2013) 043, [\href{http://arxiv.org/abs/1307.0988}{{\tt arXiv:1307.0988}}].
  [Erratum: JHEP09,180(2014)].

\bibitem{Tanabashi:2018oca}
{\bf Particle Data Group} Collaboration, M.~Tanabashi et~al., {\it {Review of
  Particle Physics}},  {\em Phys. Rev.} {\bf D98} (2018), no.~3 030001.

\bibitem{deSalas:2017kay}
P.~F. de~Salas, D.~V. Forero, C.~A. Ternes, M.~Tortola, and J.~W.~F. Valle,
  {\it {Status of neutrino oscillations 2018: 3$\sigma$ hint for normal mass
  ordering and improved CP sensitivity}},  {\em Phys. Lett.} {\bf B782} (2018)
  633--640, [\href{http://arxiv.org/abs/1708.01186}{{\tt arXiv:1708.01186}}].

\bibitem{Rodejohann:2012cf}
W.~Rodejohann and H.~Zhang, {\it {Simple two Parameter Description of Lepton
  Mixing}},  {\em Phys. Rev.} {\bf D86} (2012) 093008,
  [\href{http://arxiv.org/abs/1207.1225}{{\tt arXiv:1207.1225}}].

\bibitem{KamLAND-Zen:2016pfg}
{\bf KamLAND-Zen} Collaboration, A.~Gando et~al., {\it {Search for Majorana
  Neutrinos near the Inverted Mass Hierarchy Region with KamLAND-Zen}},  {\em
  Phys. Rev. Lett.} {\bf 117} (2016), no.~8 082503,
  [\href{http://arxiv.org/abs/1605.02889}{{\tt arXiv:1605.02889}}]. [Addendum:
  Phys. Rev. Lett.117,no.10,109903(2016)].

\bibitem{Alessandria:2011rc}
{\bf CUORE} Collaboration, F.~Alessandria et~al., {\it {Sensitivity of CUORE to
  Neutrinoless Double-Beta Decay}},  \href{http://arxiv.org/abs/1109.0494}{{\tt
  arXiv:1109.0494}}.

\bibitem{KamLANDZen:2012aa}
{\bf KamLAND-Zen} Collaboration, A.~Gando et~al., {\it {Measurement of the
  double-$\beta$ decay half-life of $^{136}Xe$ with the KamLAND-Zen
  experiment}},  {\em Phys. Rev.} {\bf C85} (2012) 045504,
  [\href{http://arxiv.org/abs/1201.4664}{{\tt arXiv:1201.4664}}].

\bibitem{Albert:2014fya}
{\bf EXO-200} Collaboration, J.~B. Albert et~al., {\it {Search for
  Majoron-emitting modes of double-beta decay of $^{136}$Xe with EXO-200}},
  {\em Phys. Rev.} {\bf D90} (2014), no.~9 092004,
  [\href{http://arxiv.org/abs/1409.6829}{{\tt arXiv:1409.6829}}].

\bibitem{Abt:2004yk}
I.~Abt et~al., {\it {A New $Ge^{76}$ Double Beta Decay Experiment at LNGS:
  Letter of Intent}},  \href{http://arxiv.org/abs/hep-ex/0404039}{{\tt
  hep-ex/0404039}}.

\bibitem{Guiseppe:2011me}
{\bf Majorana} Collaboration, C.~E. Aalseth et~al., {\it {The Majorana
  Experiment}},  {\em Nucl. Phys. Proc. Suppl.} {\bf 217} (2011) 44--46,
  [\href{http://arxiv.org/abs/1101.0119}{{\tt arXiv:1101.0119}}].

\bibitem{CarcamoHernandez:2017kra}
A.~E. C\'arcamo~Hern\'andez and H.~N. Long, {\it {A highly predictive $A_{4}$
  flavour 3-3-1 model with radiative inverse seesaw mechanism}},  {\em J.
  Phys.} {\bf G45} (2018), no.~4 045001,
  [\href{http://arxiv.org/abs/1705.05246}{{\tt arXiv:1705.05246}}].

\bibitem{Ishimori:2010au}
H.~Ishimori, T.~Kobayashi, H.~Ohki, Y.~Shimizu, H.~Okada, and M.~Tanimoto, {\it
  {Non-Abelian Discrete Symmetries in Particle Physics}},  {\em Prog. Theor.
  Phys. Suppl.} {\bf 183} (2010) 1--163,
  [\href{http://arxiv.org/abs/1003.3552}{{\tt arXiv:1003.3552}}].

\end{thebibliography}\endgroup



\end{document}